\documentclass[11pt,twoside]{article}
\usepackage{./asp2014}
\usepackage[usenames,dvipsnames]{color}

\aspSuppressVolSlug
\resetcounters

\begin{document}

\title{Science with an ngVLA: Accretion and Jets in Local Compact Objects}
\author{Deanne L. Coppejans,$^1$ James C.A. Miller-Jones,$^2$ Elmar G. K\"{o}rding,$^3$ Gregory R. Sivakoff$^4$ and Michael P. Rupen$^5$
\affil{$^1$Center for Interdisciplinary Exploration and Research in Astrophysics (CIERA) and Department of Physics and Astronomy, Northwestern University, Evanston, IL 60208; \email{deanne.coppejans@northwestern.edu}}
\affil{$^2$International Centre for Radio Astronomy Research -- Curtin University, GPO Box U1987, Perth, WA 6845, Australia; \email{james.miller-jones@curtin.edu.au}}
\affil{$^3$Department of Astrophysics/IMAPP, Radboud University, PO Box 9010, NL-6500 GL Nijmegen, the Netherlands; \email{elmar@koerding.eu}}
\affil{$^4$Department of Physics, University of Alberta, CCIS 4-183, Edmonton, AB T6G 2E1, Canada; \email{sivakoff@ualberta.ca}}
\affil{$^5$Herzberg Astronomy and Astrophysics Research Centre, 717 White Lake Road, Penticton, BC, V2A 6J9, Canada; \email{michael.rupen@nrc-cnrc.gc.ca}}}

\paperauthor{Deanne L. Coppejans}{deanne.coppejans@northwestern.edu}{}{Northwestern University}{Center for Interdisciplinary Exploration and Research in Astrophysics (CIERA) and Department of Physics and Astronomy}{Evanston}{Illinois}{60208}{United States}
\paperauthor{James C.A. Miller-Jones}{james.miller-jones@curtin.edu.au}{}{Curtin University}{International Centre for Radio Astronomy Research}{Perth}{Western Australia}{WA 6845}{Australia}
\paperauthor{Elmar G. K\"{o}rding}{elmar@koerding.eu}{}{Radboud University}{Department of Astrophysics/IMAPP}{Nijmegen}{Gelderland}{NL-6500}{Netherlands}
\paperauthor{Gregory R. Sivakoff}{sivakoff@ualberta.ca}{}{University of Alberta}{Department of Physics}{Edmonton}{Alberta}{AB T6G 2E1}{Canada}
\paperauthor{Michael P. Rupen}{michael.rupen@nrc-cnrc.gc.ca}{}{Herzberg Astronomy and Astrophysics Research Centre}{}{Penticton}{British Columbia}{V2A 6J9}{Canada}

\begin{abstract}
Despite the prevalence of jets in accreting systems and their impact on the surrounding medium, the fundamental physics of how they are launched and collimated is not fully understood. Radio observations of local compact objects, including accreting stellar mass black holes, neutron stars and white dwarfs, probe their jet emission. Coupled with multi-wavelength observations, this allows us to test the underlying accretion-outflow connection and to establish the relationship between the accretor properties and the jet power, which is necessary to accurately model jets. Compact accretors are nearby, numerous and come in a range of accretor properties, and hence are ideal probes for the underlying jet physics. Despite this there are a number of key outstanding questions regarding accretion-driven outflows in these objects that cannot be answered with current radio observations. The vastly improved sensitivity, polarization capabilities, spatial resolution and high-frequency coverage of the ngVLA will be crucial to answering these, and subsequently determining the fundamental physics behind accretion and jets at all physical scales. 
\end{abstract}

\section{Introduction}

Jets are ubiquitous, occurring in accreting systems as diverse as proto-stars and Active Galactic Nuclei (AGN). This intimate relation between accretion and directed outflow suggests that jets play an important role in re-directing angular momentum, energy, and magnetic fields from the infalling gas, while the consequent injection of mass and energy into the accretion reservoir is likely important for phenomena ranging from the initial mass function of stars to the evolution of galaxies. Despite this, we still do not fully understand the mechanism whereby jets are launched and collimated or how this is affected by simple physical properties of the accretor. One of the best ways to study the fundamental link between jets and accretion is through observations of accreting compact objects in the local universe. These accreting black holes, neutron stars and white dwarfs evolve through accretion states on observable time-scales (days to months) and are common enough that there are nearby objects for detailed study, as well as large numbers available for statistical comparisons. Furthermore, standard techniques in the field of stellar binaries can directly determine fundamental parameters such as mass and binary separation, and even constrain the evolutionary history of these systems. Through comparative studies of the different classes of compact accretors, we can determine how the properties of the accretor (mass, spin, accretion rate, absence or presence of a solid surface, strength of magnetic field, importance of relativistic effects) are related to the jet power, and can probe the underlying connection between the accretion flow and the jets. These are very complex systems, with a rich variety of emission from the accretor, its companion, the accretion flow, and the jet. In many systems, radio observations are unique in providing a view of the jet (via synchrotron emission) unconfused by all the other emission sources, and this combined with the high angular resolution of radio interferometers has made radio data indispensable to accretion/outflow studies of compact objects. In this article we will highlight some of these questions for black hole (Section \ref{sec:BH}), neutron star (Section \ref{sec:NS}) and white dwarf (Section \ref{sec:WD}) compact binaries.

Even in an era where the SKA will likely be operating, the ngVLA will play a critical role answering these questions. In addition to its greater sensitivity and northern sky coverage, ngVLA observations at greater than 10 GHz will detect jet emission closer to where jet particles are first accelerated. At these frequencies, high angular resolution imaging will expose (transient) jet ejecta and better resolve the compact, steady jet associated with sources emitting at a low fraction of their maximum (Eddington-luminosity limited) mass accretion rate. High frequency observations will also mitigate issues of confusion and scattering that might occur at lower frequencies. The large spectral lever arm allowed by ngVLA will also provide much better constraints on the radio spectra, and hence disentangle different emission mechanisms.

\section{Black hole X-ray binaries}\label{sec:BH}

Black hole X-ray binaries (BHXRBs) are ideal testbeds for studying how the mass accretion rate and accretion flow geometry (as inferred from strong X-ray emission) affect the launching of outflows (jets and winds) from compact objects. 
A direct empirical link between accretion properties (and their changes on timescales of days to months) and radio emission from the jets has been observed in numerous outbursts of some tens of BHXRBs, and holds great promise for answering fundamental questions regarding how and why accretion geometry changes, and how this affects jet properties. However, current instrumentation has limited our progress in answering these fundamental questions; more sensitive observations with broader spectral coverage are necessary.

\subsection{The accretion-outflow connection and the Fundamental Plane of Black Hole Activity}

Low-mass BHXRBs show repeated outbursts in which the accretion rate and hence the source luminosity increases by several orders of magnitude from quiescence ($L <10^{-5.5} L_{\rm Edd}$, where $L_{\rm Edd}$ is the Eddington luminosity), while the other parameters of the system (e.g.\ mass, inclination, spin) stay effectively constant. 
At luminosities up to $L\sim 10^{-1} L_{\rm Edd}$, the accretion scenario is best described by a thin disk that is likely truncated at a few tens to hundreds of gravitational radii \citep[e.g.][]{Esin97}. Within this radius is a hot, optically-thin, vertically-extended flow (such as an advection-dominated accretion flow), whose hot electrons comptonize the disk and accretion flow photons, leading to the observed power-law X-ray emission. 
During this phase of an outburst, the radio emission is strongly correlated with the X-ray luminosity (the radio/X-ray correlation, \citealt{Hannikainen1998}, see Figure \ref{Fig:plane}), interpreted as a relation between accretion rate and outflow in a compact, steady jet. 
This correlation is a major foundation upon which was built the (broadly used) Fundamental Plane of Black Hole Activity (FPBHA), which connects radio luminosity, X-ray luminosity, and black hole mass across all mass scales from BXHRBs to AGN \citep{Merloni2003,Falcke2004}.%

Around $L\sim 10^{-1} L_{\rm Edd}$, the X-ray spectrum softens rapidly as the accretion geometry transforms to a thin disk that extends all the way to the innermost stable circular orbit \citep[ISCO; see][]{Shakura73}. During this transition, the compact core radio jet is observed to be quenched \citep{Tananbaum1972} and instead one often observes bright, transient ejecta moving away from the central binary at relativistic speeds \citep[e.g.][]{Mirabel94,Tingay95}. Being well-separated from the core, the observed radio emission then no longer falls on the FPBHA. When the source moves back towards quiescence, the core radio emission reappears and the source returns to the radio/X-ray correlation (and the FPBHA).

With better coverage, many sources have deviated from the standard FPBHA, seeming instead to follow a `radio-quiet' branch of the radio/X-ray correlation at $10^{-3} \lesssim L/L_{\rm Edd} \lesssim 10^{-1}$ \citep[see, e.g.,][]{Corbel2004,Soleri2011}. This directly challenges the simple picture presented of the FPBHA with its notion of a uniform disk-jet connection in accreting black holes across all masses. The mean slope of the radio spectrum is currently our only known difference between the radio-quiet and radio-loud branches in BHXRBs \citep{Espinasse2018}. 
\citet{Meier2001} showed that systems with a thick disk and a rapidly rotating geometry (i.e., a Kerr BH as opposed to a Schwarzschild BH) produce stronger radio-emitting jets for a given accretion power, and that this supports the paradigm where radio-quiet AGN are powered by non-spinning supermassive BHs.
While this might suggest `radio-quiet' BHXRBs are powered by Schwarzschild stellar-mass BHs, several radio-quiet BHXRBs have been observed to transition back to the the main radio/X-ray correlation at $L\sim 10^{-3} L_{\rm Edd}$ (e.g.\ H1743$-$322; \citealt{Coriat2011}).
Light curves of a sample of BHXRBs over a large frequency range when $L \sim 10^{-3} L_{\rm Edd}$ (and radio luminosity is relatively low) are needed to determine if the radio-quiet track continues, or if all radio-quiet BHXRBs transition back to the the main radio/X-ray correlation. These observations will determine the nature of the radio-quiet branch and inform us on when the FPBHA is valid and what additional parameters (such as black hole spin or residual inner disk emission) might be needed.

The ngVLA can also help determine how jets switch off. The sensitivity of current radio telescopes is insufficient to determine whether the compact steady radio jet is fully quenched or only less luminous when the thin accretion disk reaches the ISCO \citep[e.g][]{Tananbaum1972,Drappeau2017}. Current observations indicate that the emission is quenched by more than 2.5 orders of magnitude relative to when the inner accretion flow is geometrically thick \citep[e.g][]{Russell2011}. Given the gain in sensitivity, ngVLA observations of a sample of BHXRBs in the soft state will probe this quenching factor to more than an order of magnitude deeper than current limits, and significantly deeper than that observed between radio-loud and radio-quiet AGN. An important caveat is that the coronal emission from the secondary star, or possibly winds, could be detectable at these limits, and the source of any detected radio emission would need to be confirmed using the improved ngVLA spectral and polarization constraints. Careful selection of targets with less active secondary stars would also help to mitigate this. If the quenching is any stronger, new suppression mechanisms (e.g. low radiative efficiency, Poynting flux dominated jet, etc.) would be required beyond the geometrical factor suggested by \citet{Meier2001}.

Currently it is not clear how BHXRBs behave in quiescence (the state in which BHXRBs and super-massive black holes spend most of their time). In particular, does the radio/X-ray correlation and FPBHA universally extend towards the lowest luminosities? Answering this will help to break the degeneracies from X-ray spectral fitting in determining the nature of the accretion flow, and will enable us to interpret the radio-bright sources seen in and towards globular clusters \citep{Strader2012,Tetarenko2016}. The ngVLA will extend the radio/X-ray correlation more than 2 orders of magnitude towards lower luminosities.

\subsection{Jet properties}

Fundamental properties of the jets in BHXRBs are currently unknown. For instance, when and how does the jet decelerate and expand as it moves away from the central black hole? What is the shape (or shapes) of jets? What is the radiative (and kinetic) power of jets? What are the magnetic field configurations of jets? And how do all of these vary as accretion flow properties (e.g.\ the accretion geometry) change? To date, studies of a handful of individual sources have allowed us to piece some of this information together, although
questions still remain, and a lack of spectral coverage, spatial resolution and insufficient sensitivity to monitor the jet over sufficiently short time-scales is preventing progress. The ngVLA will be transformative in enabling simultaneous studies of all of these questions in individual sources.

For instance, the value of high-cadence multi-wavelength observations in determining jet properties was recently demonstrated through observations of the BHXRB V404 Cygni in outburst.  Via the use of subarrays, high time resolution light curves at multiple frequencies enabled \citet{Tetarenko2017} to use physical modelling to determine some of the key jet parameters. However, V404 Cygni is a nearby source and one of the few that reached the Eddington luminosity, so higher-sensitivity radio observations are necessary to extend such studies to a sample of BHXRBs.
In particular, ngVLA observations with wide spectral coverage (potentially using subarrays) are needed to monitor variability propagating down through the frequency bands as they move down the jet, determining the jet velocity from the time delays between bands, and the opening angle from the variability timescales at different frequencies. The high spatial resolution, and great sensitivity at high frequencies will allow direct imaging of jets to much lower surface brightnesses and thus farther away from the central source, given that the transient ejecta fade due to adiabatic expansion as they move away from the core. This will track the expansion and deceleration of the jet. 

By resolving the magnetic field direction and fractional linear polarization, polarization studies can reveal the geometry of the compact steady jet and discrete jet ejects \citep[e.g.,][]{Curran2015}, even when those geometries are not directly imaged. Such studies have been strongly sensitivity limited to date, as the polarization fractions are typically of the order of a few percent in compact jets to several tens of percent in transient ejecta. The ngVLA will extend these studies to a wider (radio-fainter) range of BH and NS XRB systems, with the potential to trace jet changes to accretion changes. With long baselines, ngVLA will not only verify the use of unresolved polarimetry for tracing jet geometry, but also test whether spine-sheath jet geometries invoked in blazars  \citep[e.g.,][]{Attridge1999,Pushkarev2012} apply to XRBs.

By observing a sample of sources over the wide frequency coverage of the ngVLA, we can additionally observe how the jet break frequency depends on the properties of the accretion flow. Observations of the BHXRBs MAXI J1836-194 and MAXI J1659-152 suggest that the jet break evolves to lower radio frequencies as X-ray spectrum softens and the accretion flow geometry changes during an outburst \citep{Russell13,Russell14,vanderHorst13}. The jet-break frequency is thus expected to move through the ngVLA bands during the hard-to-soft state transition.

With its high-frequency coverage, significantly enhanced sensitivity, improved spatial resolution, and strong $uv$-coverage, ngVLA images of BHXRBs will be dramatically improved. We expect to see the sort of images that we have of extragalactic sources (e.g.\ Cyg A), revealing details of the complex interactions between the jet and the surrounding medium. This is akin to the qualitative difference that came in studies of extragalactic jets with the advent of the VLA. With the ngVLA we can make these images and study objects that evolve on human (and even undergraduate) timescales.

\section{Neutron star X-ray binaries}\label{sec:NS}

\articlefigure{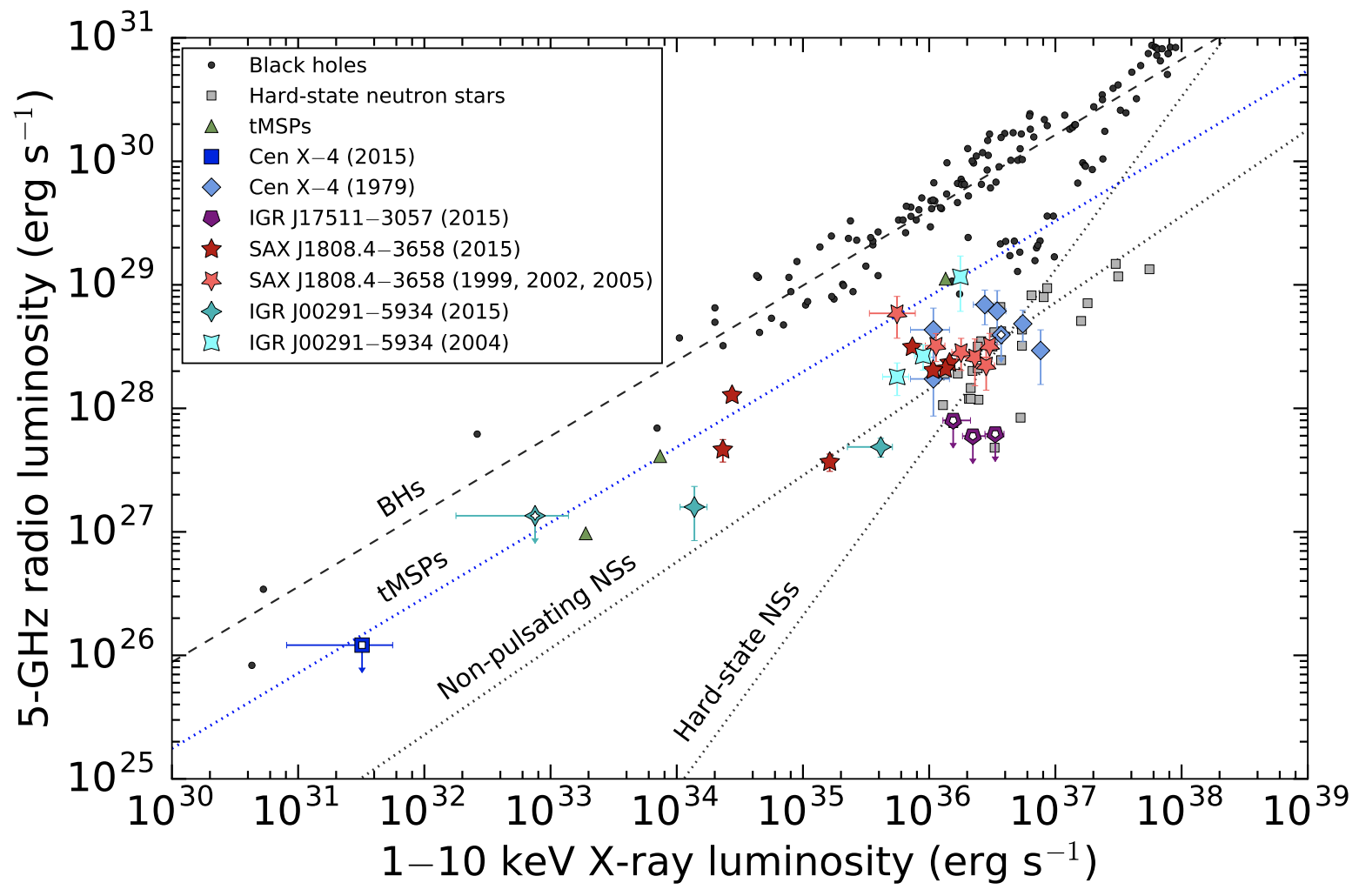}{Fig:plane}{Black hole accretors (from BHXRBs to AGN) show a relation between the radio luminosity ($\rm L_r$) and the X-ray luminosity ($\rm L_x$) of the form $\rm L_r\propto L_x^\beta$, which relates the jet power to the accretion rate. This is referred to as the radio/X-ray correlation. The question of whether neutron star systems show a similar relation remains open, although NSXRBs are radio quiet in comparison to their BHXRB counterparts \citep[e.g.][]{Fender2001, Migliari2006}. Unlike current instruments, the ngVLA will have the sensitivity to study NSXRBs down to the crucial phase space $\rm L_x\leq10^{36}\,erg\,s^{-1}$. Credit: Figure 6 from \citet{Tudor2017}.}

Neutron star X-ray binaries (NSXRBs) are compact binaries with a neutron star accretor. In contrast to the BHXRBs, the accretor has a surface and a stellar magnetic field. It is currently unknown how these properties directly affect jet formation and collimation, so comparative studies between these two classes of object are a powerful tool to determine this. At present these studies are severely hampered by a lack of sensitivity, as NSXRBs are fainter than their black hole counterparts at radio wavelengths. Increased sensitivity with higher frequency coverage and better spatial resolution are necessary to determine which classes of NSXRBs launch jets (and hence which system parameters enable/prevent jet launching) and how the accretion-outflow connection differs from the BHXRBs.

In the transitional millisecond pulsars (tMSPs; a class of rapidly-rotating neutron stars that switch between accretion and rotation powered phases during the last stages of recycling), it is currently not known whether the observed radio emission is produced by a jet or another mechanism. In at least one tMSP, anti-correlated radio and X-ray emission has been detected that cannot readily be explained by traditional accretion-outflow models \citep{Bogdanov2017}. The key to determining the source of the radio emission is simultaneous X-ray and ngVLA radio observations of a sample of tMSPs with sufficient sensitivity to track the spectral evolution over the radio variability time-scales ($\sim10\,$s). At 8 GHz, the ngVLA can reach the sensitivities required to track the in-band spectral evolution of the observed $\sim200\,\mu$Jy flares in less than a second\footnote{assuming a $0.21\,\mu$Jy/beam RMS for a 1 hour observation using the full bandwidth (ngVLA memo \#17)} and effectively sample the flaring emission.

Similarly, it is not known whether strong magnetic fields ($>10^{12}$ G) inhibit the formation of jets. The lack of observed jets in strongly magnetic NSXRBs \citep[e.g][]{Fender2000} has resulted in a number of jet formation models that suppress jet formation at high magnetic field strengths \citep[e.g.][]{Massi2008}. Recently however, radio detections of a few such systems have been found that are consistent with jet synchrotron emission \citep{vandenEijnden17,vandenEijnden2018}. These detections of nearby systems are too faint (10--100 $\mu$Jy) to distinguish jet emission from other mechanisms. Other possibilities include emission from shocks in the interaction between the accretion flow and magnetosphere or a propeller-driven outflow, but for Her X-1, \citeauthor{vandenEijnden17} state that these two scenarios are unlikely. Stronger constraints on the linear and circular polarization fractions from the improved sensitivity of the ngVLA will be important in determining the relevant emission mechanism, as will the improved spectral resolution and access to higher frequencies to search for characteristic spectral features such as jet breaks.

The key to comparing the accretion-outflow connection in BHXRBs and NSXRBs is through the radio/X-ray plane (see Figure \ref{Fig:plane}) via measurements of the radio ($\rm L_r$) and X-ray luminosity ($\rm L_x$). As described in Section \ref{sec:BH}, this gives the relationship between the jet power and accretion rate. At present it is unclear whether the different classes or states of neutron star show a single behavior. Radio measurements of systems with $\rm L_x\leq10^{36}\,erg\,s^{-1}$ are necessary to determine the relation/s, but there are very few detections in this range. The ngVLA sensitivity is necessary to effectively sample this phase-space, as the NSXRBs are comparatively radio-quiet to their BH counterparts. \citet{Gallo2018} found that the BH systems are a factor $\sim22$ more radio-loud than the NSXRBs for a given x-ray luminosity. Clarifying this NSXRB behavior on the radio/x-ray plane will make it possible to determine what role the radiative efficiency of the accretion plays in NSXRBs and BHXRBs. 

\section{Cataclysmic Variables}\label{sec:WD}

Cataclysmic Variables (CVs) are binary stars in which a white dwarf accretes material from a low-mass main-sequence star via Roche-lobe overflow \citep[see][]{Warner1995}. In the `non-magnetic' CVs (the subject of this section and hereafter referred to simply as `CVs'), the transfered material accretes via an accretion disk, as the magnetic field strength of the white dwarf (at $\lesssim10^6$ G) is not sufficient to channel the accretion flow from the disk. Until recently, it was believed that CVs were one of the few classes of compact accretors to not launch jets and they have consequently been used to constrain jet-launching models in compact accretors \citep[e.g][]{Soker2004}. Other classes of accreting white dwarfs such as the novae \citep[e.g.][]{Sokoloski2008} and symbiotic stars \citep[e.g.][]{Brocksopp2004} are known to launch jets, but CVs were thought to be the exception. In 2008 however, \citeauthor{Koerding2008} showed that the same accretion-outflow cycle observed in XRBs could be mapped to CVs. This and subsequent studies \citep{Miller-Jones2011,Russell2016,Mooley2017} indicated a likely jet in the prototypical CV SS Cyg. If CVs are launching jets this will have significant implications for jet-launching models in compact accretors because CVs were previously used to constrain these models. They are numerous, nearby, comparatively better understood than the black hole and neutron star accretors and are non-relativistic. Additionally, the Gaia survey \citep{GaiaCollaboration2016} has published the distances to many nearby CVs, making it is possible to convert observables into physical parameters and do comparative studies between systems. For all these reasons, if CVs are launching jets, they will be important probes of accretion and jet physics in compact accretors. The enhanced sensitivity, higher frequency coverage, and higher resolution of the ngVLA are necessary to establish whether this is the case.

CVs are faint at radio frequencies (on the order of 10s to 100s of $\mu$Jy at 300 pc), and only with the recent VLA upgrade have we had the sensitivity required to detect even the handful of very nearby systems \citep{Koerding2011,Coppejans2015,Coppejans2016}. With the notable exception of the extraordinarily bright CV SS Cyg, the sensitivity of current instruments is insufficient to determine the spectral indices, put strong constraints on the polarization fraction or determine the spatial extent of the emitting region (through VLBI observations) of CVs.

There are five key parts to the question of whether CVs launch jets that require ngVLA observations to make progress. \textit{First}, we need to determine the broad spectral behavior to ascertain the emission mechanism. \textit{Second}, the radio spectrum needs to be observed out to 116\,GHz to look for a possible jet break (analogously to the XRBs). \textit{Third}, observations mapping the spectral evolution, timing and polarization properties of the $\sim$minute time-scale flares equivalent to those of SS Cyg \citep{Mooley2017} are needed for a sample of CVs. This is necessary to be able to distinguish jet emission from alternative mechanisms such as magnetic reconnections in the disk \citep[see][]{Meintjes2016}. Importantly, current observations show different polarization behavior during the flares of the `magnetic' and `non-magnetic' CVs, indicating different radio emission mechanisms in these classes. In the `magnetic CVs'\footnote{CVs where the magnetic field strength of the white dwarf exceeds $\approx10^6$ G and truncates the disk at the Al\'{v}en radius. The strongest field systems prohibit the formation of a disk.}, highly circularly polarized flaring emission characteristic of electron-cyclotron maser emission has been detected in some systems, while in others the low polarization limits are more indicative of gyrosynchrotron emission \citep[][]{Abada-Simon1993, Dulk1983, Chanmugam1987, Wright1988, Pavelin1994, Meintjes2005, Mason2007, Barrett2017}. In contrast only one `non-magnetic CV' has shown a highly circularly polarized flare \citep{Coppejans2015}. Deep ngVLA limits on the polarization fractions of the flares will help to reveal the impact of the magnetic field strength of the white dwarf on the emission and consequently determine the radio emission mechanisms at play in these objects. \textit{Fourth}, to conclusively determine if CVs launch jets, observations that spatially resolve the jet emission are necessary. The resolution at 8 GHz (assuming natural weighting) is 49 mas (ngVLA memo \#17), which will probe physical scales on the order of $10^{14}$ cm (1000 times the orbital separation) at 300 pc. VLBA observations of SS Cyg have shown indications of extended emission at a 4$\sigma$ level \citep{Russell2016}, but the majority of the CV population are too faint for comparable observations at present. \textit{Fifth}, a larger sample of detected CVs are necessary to sample the full range of CV properties (e.g. accretor mass, accretion rate and magnetic field strength) to determine how these properties affect jet-launching. Assuming a 0.21 $\mu$Jy RMS for a 1 hour observation at 8 GHz (ngVLA memo \#17), we will detect emission from even the faintest observed CVs at a 5$\sigma$ level out to $\sim1700$ pc. Using a lower-limit on the space density  of CVs of $\rm2\times10^{-6}\,pc^{-3}$ \citep{Pretorius2012}, this puts a lower-limit on the number of CVs that can be detected at $\gtrsim$7000. This is a minimum of a factor 700 enlargement of our current sample, which will enable conclusive statistical studies. Given the detection volume of CVs, and the large ngVLA field-of-view particularly at lower frequencies, CVs will also be detected in ngVLA survey data and transient searches using multiple beams. Unlike with current instruments, these systems will be detected serendipitously and will not exclusively require targeted detection experiments.

We end this section on a related note, by briefly highlighting two white dwarf binaries (AE Aqr and AR Sco) for which radio observations are driving significant advances in our understanding of the physics of matter transfer onto rapidly rotating, strongly magnetic compact stars; in both cases the ngVLA will significantly further progress. AE Aqr is a CV with a white dwarf that is highly magnetic and has a short spin period such that the spinning magnetosphere causes the majority of the transfered material to be flung out of the system rather than accreted \citep[a propeller phase;][]{Wynn1997,Meintjes2000,Meintjes2005}. AR Sco consists of a white dwarf and M-dwarf binary in a short orbital period (3.55 h), but unlike CVs there is no evidence of accretion. It has been called the first (and to date, only) white dwarf pulsar as it is powered by the spin-down of the white dwarf rather than via accretion (\citealt{Marsh2016,Buckley2017,Stiller2018}, see however \citealt{Potter2018}). Studies of AE Aqr and AR Sco have had important implications for our understanding of the evolutionary stages of mass accretion in white dwarfs, and the modelling of these mass transfer stages is likely relevant to the higher mass compact binaries such as the accreting millisecond pulsars \citep{Campana2018}. For both of these objects, the improved sensitivity of the ngVLA will enable higher time- and spectral resolution studies with finer sampling of the polarized flares, for more detailed modelling of the radio emission. Another promising avenue is the serendipitous discoveries of additional objects like these for more context on these processes.

\section{Conclusion}

ngVLA multi-band observations will track the spectral, spatial, polarization and variability properties of jets and outflows in compact accretors on scales and luminosities that are currently unreachable. To probe the accretion-ejection connection, complementary optical, UV and X-ray observations are additionally necessary to monitor the accretion flow. In the XRBs, this will be accomplished through simultaneous X-ray observations. In the CVs, emission from the accretion flow peaks at optical, UV and X-ray wavelengths (in order of decreasing distance to the white dwarf). Current, planned and possible missions that can provide this complementary coverage of the accretion flow include Evryscope, LSST, WFIRST, LUVOIR, Athena, STROBE-X, the successor to TESS (possibly FINESSE), and the Lynx X-ray Surveyor. The vast improvement of radio observations offered by the ngVLA, as well as this complementary coverage from multi-wavelength future missions, are key to answering the outstanding questions related to accretion and jet physics in compact accretors, and by extension other classes of accretors. 

\clearpage 

\acknowledgements
We thank the referee for comments and suggestions that have significantly improved this paper. GRS acknowledges support by an NSERC Discovery Grant. JCAM-J is the recipient of an Australian Research Council Future Fellowship (FT140101082).

\bibliography{ngVLA_jets}


\end{document}